\documentclass[aps,11pt]{revtex4}
\usepackage[dvips]{graphicx}

\usepackage{amsmath}
\usepackage{bm}

\begin{document}

\preprint{}

\title{How large could the R-parity violating couplings be?}
\author{Pavel Fileviez P\'erez}
\affiliation{Pontificia Universidad Cat\'olica de Chile \\
Facultad de F{\'\i}sica, Casilla 306 \\
Santiago 22, Chile.}
\begin{abstract}
We investigate in detail the predictions coming from the $d=4$
operators for proton decay. We find the most general constraints 
for the $R$-parity violating couplings coming from proton decay, 
taking into account all fermion mixing and in different supersymmetric
scenarios. 
\end{abstract}
\maketitle

\section{Introduction}

The minimal supersymmetric extension of the standard model (MSSM) is
one of the most popular candidates for the physics beyond the standard 
model. It is well known that in general it is possible to write down
in the superpotential terms which violate the Baryon($B$) or Lepton($L$)
number, giving us the possibility to describe the neutrino mass through
those interactions. For several phenomenological aspects of $R$ parity
violating interactions see reference~\cite{review}. 

Due to the presence of the $B$ violating interactions in the MSSM 
superpotential it is very important to understand the 
constraints coming from the proton decay searches.  
The decay of the proton, which was predicted long ago by 
Pati and Salam~\cite{PatiSalam}, is one the most important 
constraints for physics beyond the standard model, 
particularly for grand unified theories~\cite{GUTs}.
 
It is well known that in supersymmetric scenarios the $d=4$ and $d=5$
contributions to the decay of the proton are the most important, while 
in non-supersymmetric scenarios the gauge $d=6$ operators are the
dominant. Assuming that the so-called $R$-parity is an exact symmetry 
of our model the dangerous $d=4$ operators are forbidden and there is the
possibility to describe the cold dark matter in the Universe, since
the lightest supersymmetric particle (Neutralinos) will be
stable. Those are the main reasons to impose the $R$ parity in
the context of the minimal supersymmetric standard model. 
See also reference~\cite{Goran} for the possibility to have 
$R$-parity as an exact symmetry coming from grand unified theories. 

In general, it is very difficult to satisfy the proton decay 
experimental bounds in the context of minimal models. For example 
to know about the status of the minimal supersymmetric $SU(5)$ model
see reference~\cite{ProtonSU(5)}. However, in realistic models there
is more freedom, and it is still possible to satisfy all bounds. 
New experimental bounds~\cite{experiments} are welcome, in order 
to have better constraints and if proton decay is found we will able 
to test the idea of grand unified theories.  
We understand all contributions to the decay of the proton, however it
is very difficult to know how realize the test of a given model. 
Recently, it has been pointed out the possibility to make a 
clear test of grand unified theories with symmetric Yukawa couplings 
through the proton decay into antineutrinos~\cite{Pavel}. In the context of the
minimal renormalizable supersymmetric flipped SU(5) model it has been
pointed out the possibility to use also the ratio between the decays 
into charged leptons in order to make a clear test~\cite{Ilja1}. 
At the same time there is a very interesting possibility to 
rotate proton decay away in the context of flipped $SU(5)$
models~\cite{Ilja2}. A second crucial issue about proton decay is the possibility to find
an upper bound on the total proton lifetime, since in this case we 
will hope that proton decay will be found. Recently, it has been 
found an very conservative upper bound on the lifetime 
of the proton~\cite{Ilja3}, therefore there is hope to test all those
ideas if the decay of the proton is found.

The constraints for the R-parity violating couplings coming 
from proton decay in low energy supersymmetry have been studied 
long ago~\cite{ProtonR}. However, up to now it has not been
found the most general constraints. In this Letter 
our main task is study the predictions coming from the
$d=4$ operators in the context of the MSSM. In order to find the 
most general constraints for the relevant couplings we take into 
account all flavour mixing, and compute the amplitudes for the 
different channels using the Chiral Lagrangian Techniques. 
We compare the constraints for the R-parity violating
couplings in the context of the minimal supersymmetric 
standard model in two scenarios, in low energy SUSY (See for example~\cite{Haber}) 
and in SPLIT supersymmetry~\cite{Arkani}, a new interesting scenario where all
scalars, except for one Higgs, are very heavy and the fermions are
light. In this scenario it is possible to achieve the unification of gauge
couplings and we have a natural candidate for the cold dark matter in
the Universe if the $R$-parity is imposed (See also~\cite{Gupta} for a
possibility to have the Neutralino as cold dark matter candidate even
if the $R$-parity is broken). See references~\cite{SPLIT}
for different phenomenological studies in this scenario.
Notice that since there is not a theory for $R$-parity violation, the
only thing that we can do is find the constraints for these couplings,
and the question about the possibility to test those theories through
proton decay remains open.  
%
%
\section{R-parity violation and the decay of the proton}
%
In the minimal supersymmetric standard model we are allowed to write the
following terms in the superpotential which violate the so-called $R$
parity:
\begin{equation}
W_{NR} = \alpha_{ijk} \ {\hat Q}_i \ { \hat L}_j \ {\hat D}^C_k \ + \
\beta_{ijk} \ {\hat U}^C_i \
{\hat D}^C_j \ {\hat D}_k^C \ + \ \gamma_{ijk} \ {\hat L}_i \ {\hat
  L}_j \ 
{\hat E}_k^C \ + \ a_i \ {\hat L}_i \ {\hat H}
\end{equation}
In the above equation we use the usual notation for all MSSM
superfields (See for example~\cite{Haber}). 
The $R$-parity is defined as $R=(-1)^{2S} M$, where $S$ is the spin
and $M=(-1)^{3(B-L)}$ is the Matter parity, 
which is $-1$ for all matter superfields and $+1$ for Higgs and 
Gauge superfields. The coefficient $\beta_{ijk}=- \beta_{ikj}$ and 
$\gamma_{ijk}= - \gamma_{jik}$. Notice that in the above equation 
the second term violates the Baryon number, while the rest of 
the interactions violate the Leptonic number.  
It is well known that from the first and second terms we can write at
tree level the contributions to proton decay mediated by the 
${\tilde d}^C_k$ squarks. These are the most important contributions,
and we will use those to understand the constraints. 
Now let us write the decay rate of the proton using the
$R$-parity violating interactions. Using the Chiral Lagrangians and
writing all interactions in the physical basis the decay
rates of decays of the proton into charged leptons are given by:
\begin{eqnarray}
\Gamma(p \to \pi^0 e^{+}_{\beta}) &=& \frac{m_p}{64 \pi f_{\pi}^2} 
\ A^2_L \ |\alpha|^2 \ (1 + D + F)^2  \ |c(e_{\beta}^+, d^C)|^2 \\
\Gamma(p \to K^0 e^{+}_{\beta}) &=& \frac{(m_p^2 - m_K^2)^2}{32\pi
  f_{\pi}^2 m_p^3} \ A^2_L \ |\alpha|^2 \ [ 1 + \frac{m_p}{m_B} (D -F)]^2 
\ |c(e_{\beta}^+,s^C)|^2 
\end{eqnarray}
with:
\begin{eqnarray}
c(e_{\beta}^+, d^C_{\alpha})&=& \sum_{m=1}^3 \frac{(\Lambda_3^{\alpha m})^* 
\Lambda_1^{\beta m}}{m_{\tilde{d}^C_m}^2}  
\end{eqnarray}
where $\alpha,\beta=1,2$. $D$ and $F$ are the parameters of 
the chiral lagrangian, $\alpha$ is the matrix element, and $A_L$ 
takes into account the renormalization effects from $M_Z$ to 
$1$ GeV. In the case of the decay channels into 
antineutrinos, the decay rates read as:

\begin{eqnarray}
\Gamma(p \to K^+\bar{\nu})
&=& \frac{(m_p^2-m_K^2)^2}{32\pi m_p^3 f_{\pi}^2} A_L^2
\left|\alpha\right|^2 \nonumber\\
&&\times \sum_{i=1}^3 \left|\frac{2m_p}{3m_B}D \ c(\nu_i, d, s^C)
+ [1+\frac{m_p}{3m_B}(D+3F)] c(\nu_i,s, d^C)\right|^2 \\
\Gamma(p \to \pi^+\bar{\nu})
&=&\frac{m_p}{32\pi f_{\pi}^2}  A_L^2 \left|\alpha
\right|^2 (1+D+F)^2
\sum_{i=1}^3 \left| c(\nu_i, d, d^C) \right|^2
\end{eqnarray}
where:
\begin{eqnarray}
c(\nu_l, d_{\alpha}, d^C_{\beta})&=& \sum_{m=1}^3 
\frac{(\Lambda_3^{\beta m})^* 
\Lambda_2^{\alpha l m}}{m_{\tilde{d}^C_m}^2}  
\end{eqnarray}
In the above equations the 
couplings $\Lambda_1$, $\Lambda_2$ and $\Lambda_3$ are given by:
\begin{eqnarray}
\Lambda_1^{\alpha m} &=& \alpha_{ijk} \ U^{1i} \ E^{j \alpha} \ {\tilde
  D}_C^{km}\\
\Lambda_2^{\alpha l m} &=& \alpha_{ijk} \ D^{\alpha i} \ N^{jl} 
\ {\tilde  D}_C^{km}\\
\Lambda_3^{\alpha m} &=& 2 \beta_{ijk} \ U_C^{i1} \ D_C^{j \alpha} 
\ {\tilde  D}_C^{km} 
\end{eqnarray}
The mixing matrices in the above expression diagonalize the Yukawa
matrices in the following way:

\begin{eqnarray}
Y^{diag}_U &=& U^{T}_C \ Y_U \ U\\
Y^{diag}_D &=& D_C^T \ Y_D \ D\\
Y^{diag}_E &=& E_C^T \ Y_E \ E\\
Y^{diag}_N &=& N^T \ Y_N \ N \ \text{(in the Majorana case)}\\ 
Y^{diag}_N &=& N^T_C \ Y_N \ N \ \text{(in the Dirac case)}
\end{eqnarray}
Now, in order to get the constraints for the $R$-parity violating
couplings we have to use the decay channels, which could give us 
the information that we are looking for. 
From the different equations listed above for the decay rates, it is
easy to see that we have to use the decays into charged leptons and
mesons, since we cannot get new information from the
rest of the channels. In our opinion, it is the best way to get 
the constraints.
Using $m_p=938.3$ MeV, $D=0.81$, $F=0.44$, $M_B=1150$ MeV,
$f_{\pi}=139$ MeV, $\alpha=0.003 GeV^3$, $A_L=1.43$ and 
the experimental constraints~\cite{PDG} we get:
\begin{eqnarray}
|c(e^{+}, d^C)|&<& 7.6 \times 10^{-31}\\
|c(\mu^{+}, d^C)|&<& 1.4 \times 10^{-30}\\
|c(e^{+}, s^C)|&<& 4.2 \times 10^{-30}\\
|c(\mu^{+}, s^C)|&<& 4.7 \times 10^{-30} 
\end{eqnarray}
Now, assuming that all squarks have the same mass ${\tilde m}$, the
quantity $(\lambda_3^{\alpha m})^* \lambda_1^{\beta m}$ have to satisfy
the following relations:
\begin{eqnarray}
|(\lambda_3^{1m})^* \lambda_1^{1m}|&<& 3.8 \times 10^{-31} \ {\tilde m}^2 \\
|(\lambda_3^{1m})^* \lambda_1^{2m}|&<& 7.0 \times 10^{-31}  \ {\tilde m}^2 \\
|(\lambda_3^{2m})^* \lambda_1^{1m}|&<& 2.1 \times 10^{-30} \ {\tilde m}^2\\
|(\lambda_3^{2m})^* \lambda_1^{2m}|&<& 2.3 \times 10^{-30} \ {\tilde m}^2
\end{eqnarray}
where:
\begin{eqnarray}
(\lambda_3^{\alpha m})^* \lambda_1^{\beta m} &=& 
\beta_{ijk}^* \ \alpha_{lpk} \ (U^{1i}_C)^* \ (D_C^{j \alpha})^* \
  U^{1l} \ E^{p\beta}  
\end{eqnarray}
As you can appreciate the constraints for $\alpha_{ijk}$ and
$\beta_{ijk}$ are quite model dependent i.e., depend on the model for
fermion mass that we choose. Notice that we can choose for example the
basis where the charged leptons and down quarks are diagonal, however
still $U_C$ will remain, and $U=K_1 V_{CKM}^{\dagger} K_2$ (It is well known that the
proton decay predictions depend on the fermion mass model). $K_1$ and
$K_2$ are diagonal matrices containing three and two CP-violating
phases, respectively.  
Now, let us see the constraints for different values of the scalar
mass. In Table I we show the different constraints in two
supersymmetric scenarios, in low energy supersymmetry 
${\tilde m} = 10^3$ GeV and in split supersymmetry for 
the case ${\tilde m} = 10^{14}$ GeV, respectively. 
Notice that the value ${\tilde m} = 10^{14}$ GeV is basically the upper bound for the
scalar masses coming from the gluino decay~\cite{Arkani}.

\begin{table}
\caption{\label{tab:table2}Upper bounds for the R-parity violating couplings}
\begin{ruledtabular}
\begin{tabular}{lcccc}
\textbf{Couplings}&\textbf{Low energy SUSY}&\textbf{SPLIT
  SUSY}(${\tilde m}=10^{14}$ GeV)\\ 
\hline 
$|(\lambda_3^{1m})^* \lambda_1^{1m}|$ & $ 3.8 \times 10^{-25}$ & 0.0038\\
$|(\lambda_3^{1m})^* \lambda_1^{2m}|$ & $ 7.0 \times 10^{-25}$ & 0.0070\\
$|(\lambda_3^{2m})^* \lambda_1^{1m}|$ & $ 2.1 \times 10^{-24}$ & 0.0210\\
$|(\lambda_3^{2m})^* \lambda_1^{2m}|$ & $ 2.3 \times 10^{-24}$ & 0.0234\\ 
\end{tabular}
\end{ruledtabular}
\end{table}

Therefore as we can appreciate from these results the $R$-parity 
violating couplings could be large in supersymmetric scenarios with
large susy breaking scale (Split Supersymmetry). In the case of low
energy SUSY, since those couplings are quite small we usually believe
that it is a hint to believe that the $R$-parity has to be a symmetry of
our theory. However, in general this question is open.                
Now, What are the phenomenological implications coming from those constraints?

Those values are particularly interesting in order to study the implications 
in collider physics and cosmology. For example in the description of
neutrino masses, the three body decays of neutralinos and gluinos. 
As we know one the motivations of the supersymmetric scenarios with
large scalar masses is the possibility to describe the cold dark
matter in the Universe with neutralinos, since there are light and
stable if $R$-parity is impose. Now, here we study the implication of
$R$-parity violation for proton decay in those scenarios. It has been
shown in reference~\cite{Gupta} that if $R$-parity is broken in SPLIT
SUSY scenarios still the neutralinos could be a good candidate for
dark matter. The reason is the following, if we neglect the last term
in equation~$(1)$ the neutralinos can decay only into three
particles. In these scenarios the three bodies decays are suppressed
by the large scalar masses, then they can have a lifetime of the
order of the age of the Universe. Notice that in their analysis they
use large values for the $R$-parity violating couplings which are in
agreement with the constraints coming from proton decay obtained by us.     
\\
It is very interesting to know how could be the constraints for those
couplings in the context of grand unified theories. 
In the simplest unified theory, the minimal SUSY $SU(5)$, 
we can get the $R$-parity violating interactions present in the 
MSSM from the terms, $\Lambda^{ijk} \ 10_i \ \bar{5}_j \ \bar{5}_k$, $ b_i \ \bar{5}_i
\ 5_H$ and $c_i \ \bar{5}_i \ 24_H \ 5_H$. In this case at the GUT scale the
couplings satisfy the relations $\frac{\alpha_{ijk}}{2} = \beta_{ijk} = \gamma_{ijk} =
\Lambda_{ijk}= -\Lambda_{ikj}$. Taking into account these relations 
we can reduce the number of free parameters, and we could have 
better constraints. However, in our paper we only try to understand
the constraints in the context of the minimal supersymmetric standard model.
Notice that in grand unified theories with split supersymmetry the most 
important contributions for the decay of the proton are the gauge
$d=6$ contributions if the R-parity is conserved. Since here we 
are working in the context of the MSSM, we assumed that the $d=6$ 
non-renormalizable operators are not present. 
\\
As you can appreciate, we have found the most general constraints for
the $R$-parity violating couplings coming from proton decay, taking
into account all important effects. 
We hope that those results will be useful to study all 
phenomenological and cosmological implications of those interactions.    

\section{Summary}

We have investigated the constraints for the $R$-parity violating
couplings coming from proton decay in different supersymmetric
scenarios. Taking into account all fermion mixing, using the Chiral
Lagrangian Techniques and imposing the experimental bounds it has
been shown how large those couplings could be in the case of split
supersymmetry. We hope that these results will useful to discover 
SPLIT SUSY at future experiments.  

\begin{acknowledgments}
I thank Marco Aurelio Diaz, Ilja Dorsner, and Goran Senjanovi\'c 
for important discussions and comments. 
This work has been supported by CONICYT/FONDECYT 
under contract $N^{\underline 0} \ 3050068$.
\end{acknowledgments}



\begin{thebibliography}{99}
\bibitem{review} 
B.~Allanach {\it et al.}  [R parity Working Group Collaboration],
arXiv:hep-ph/9906224; R.~Barbier {\it et al.},
arXiv:hep-ph/0406039; H.~K.~Dreiner,
arXiv:hep-ph/9707435.

\bibitem{PatiSalam} J.~C.~Pati and A.~Salam,
Phys.\ Rev.\ D {\bf 8} (1973) 1240; Phys.\ Rev.\ Lett.\  {\bf 31} (1973) 661.

\bibitem{GUTs}
H.~Georgi and S.~L.~Glashow,
Phys.\ Rev.\ Lett.\  {\bf 32}, 438 (1974); 
S.~Dimopoulos and H.~Georgi, Nucl.\ Phys.\ B {\bf 193} (1981) 150; 
H. Georgi, in \textit{Particles and Fields}, ed. C. E. Carlson (AIP,
New York, 1975) 575; 
H.~Fritzsch and P.~Minkowski, Annals Phys.\  {\bf 93} (1975) 193.

\bibitem{Goran} 
C.~S.~Aulakh, B.~Bajc, A.~Melfo, A.~Rasin and G.~Senjanovic,
Nucl.\ Phys.\ B {\bf 597} (2001) 89
[arXiv:hep-ph/0004031]; 
C.~S.~Aulakh, B.~Bajc, A.~Melfo, G.~Senjanovic and F.~Vissani,
Phys.\ Lett.\ B {\bf 588} (2004) 196
[arXiv:hep-ph/0306242];
C.~S.~Aulakh and A.~Girdhar,
arXiv:hep-ph/0405074.

\bibitem{ProtonSU(5)}
T.~Goto and T.~Nihei, Phys.\ Rev.\ D {\bf 59} (1999)
115009; 
H.~Murayama and A.~Pierce, Phys.\ Rev.\ D {\bf 65}
(2002) 055009; 
B.~Bajc, P.~Fileviez~P\'erez and G.~Senjanovi\'c,
Phys.\ Rev.\ D {\bf 66} (2002) 075005; 
B.~Bajc, P.~Fileviez~P\'erez and G.~Senjanovi\'c,
hep-ph/0210374; 
D.~Emmanuel-Costa and S.~Wiesenfeldt, Nucl.\ Phys.\ B {\bf 661} (2003) 62.

\bibitem{experiments} See for example: 
C.~K.~Jung,
arXiv:hep-ex/0005046; A.~Rubbia,
arXiv:hep-ph/0407297.

\bibitem{Pavel} 
P.~Fileviez~P\'erez, Phys.\ Lett.\ B {\bf 595} (2004) 476.

\bibitem{Ilja1} I.~Dorsner and P.~Fileviez Perez,
  Phys.\ Lett.\ B {\bf 605} (2005) 391
  [arXiv:hep-ph/0409095].

\bibitem{Ilja2} I.~Dorsner and P.~Fileviez Perez,
  Phys.\ Lett.\ B {\bf 606} (2005) 367
  [arXiv:hep-ph/0409190].

\bibitem{Ilja3} I.~Dorsner and P.~Fileviez P\'erez, arXiv:hep-ph/0410198.

\bibitem{ProtonR}
L.~J.~Hall and M.~Suzuki,
Nucl.\ Phys.\ B {\bf 231} (1984) 419; 
R.~Barbieri and A.~Masiero,
Nucl.\ Phys.\ B {\bf 267} (1986) 679; V.~Ben-Hamo and Y.~Nir,
Phys.\ Lett.\ B {\bf 339} (1994) 77
[arXiv:hep-ph/9408315]; I.~Hinchliffe and T.~Kaeding,
Phys.\ Rev.\ D {\bf 47} (1993) 279; 
A.~Y.~Smirnov and F.~Vissani,
Phys.\ Lett.\ B {\bf 380} (1996) 317
[arXiv:hep-ph/9601387]; A.~Y.~Smirnov and F.~Vissani,
Nucl.\ Phys.\ B {\bf 460} (1996) 37
[arXiv:hep-ph/9506416]; G.~Bhattacharyya and P.~B.~Pal,
Phys.\ Lett.\ B {\bf 439} (1998) 81
[arXiv:hep-ph/9806214]; G.~Bhattacharyya and P.~B.~Pal,
Phys.\ Rev.\ D {\bf 59} (1999) 097701
[arXiv:hep-ph/9809493];
   
\bibitem{Haber} For a review see: 
H.~E.~Haber and G.~L.~Kane,
Phys.\ Rept.\  {\bf 117} (1985) 75.

\bibitem{Arkani}
N.~Arkani-Hamed and S.~Dimopoulos,
arXiv:hep-th/0405159.

\bibitem{Gupta}
S.~K.~Gupta, P.~Konar and B.~Mukhopadhyaya,
Phys.\ Lett.\ B {\bf 606} (2005) 384 [arXiv:hep-ph/0408296].

\bibitem{SPLIT}
G.~F.~Giudice and A.~Romanino,
Nucl.\ Phys.\ B {\bf 699} (2004) 65
[arXiv:hep-ph/0406088]; N.~Arkani-Hamed, S.~Dimopoulos, G.~F.~Giudice and A.~Romanino,
arXiv:hep-ph/0409232; A.~Arvanitaki, C.~Davis, P.~W.~Graham and J.~G.~Wacker,
arXiv:hep-ph/0406034; A.~Pierce,
Phys.\ Rev.\ D {\bf 70} (2004) 075006
[arXiv:hep-ph/0406144]; S.~h.~Zhu,
Phys.\ Lett.\ B {\bf 604} (2004) 207
[arXiv:hep-ph/0407072];
B.~Mukhopadhyaya and S.~SenGupta,
arXiv:hep-th/0407225; R.~Mahbubani,
arXiv:hep-ph/0408096; M.~Binger,
arXiv:hep-ph/0408240; J.~L.~Hewett, B.~Lillie, M.~Masip and T.~G.~Rizzo,
JHEP {\bf 0409} (2004) 070
[arXiv:hep-ph/0408248]; L.~Anchordoqui, H.~Goldberg and C.~Nunez,
arXiv:hep-ph/0408284; I.~Antoniadis and S.~Dimopoulos,
arXiv:hep-th/0411032; B.~Bajc and G.~Senjanovic,
arXiv:hep-ph/0411193; B.~Kors and P.~Nath,
arXiv:hep-th/0411201; 
M.~Aurelio~Diaz and P.~Fileviez~P\'erez,
arXiv:hep-ph/0412066; E.~J.~Chun and S.~C.~Park,
arXiv:hep-ph/0410242; K.~Cheung and W.~Y.~Keung,
arXiv:hep-ph/0408335; A.~Masiero, S.~Profumo and P.~Ullio,
arXiv:hep-ph/0412058; C.~Kokorelis,
arXiv:hep-th/0406258.

\bibitem{PDG} S. Eidelman et al, Phys. Lett. B 592, 1 (2004)


\end{thebibliography}
\end{document}